\documentclass[prl,twocolumn,showpacs]{revtex4}
\usepackage{amsmath,amssymb,graphicx,color,bm,epstopdf}

\begin{document}
\title{Proposal for a Detector of Photons with Zero Projection of Spin}

\author{I. G. Savenko}

\affiliation{COMP Centre of Excellence at the Department of Applied Physics and Low Temperature Laboratory (OVLL), Aalto University School of Science, P.O. Box 13500, FI-00076 Aalto, Finland}

\date{\today}

\begin{abstract}
We suggest an indirect method of detection of photons with zero projection of spin mediated by emission of terahertz photons. This terahertz source is based on a system of microcavity exciton polaritons in the regime of polariton BEC formation when the cavity photons acquire an effective mass being localised in the cavity and therefore receive the third spin degree of freedom (corresponding to the longitudinal polarization with chirality $\lambda=0$). The optical transitions can occur between two polariton ground states based on the light-hole and heavy-hole excitons, respectively, accompanied by the emission of terahertz radiation with controllable characteristics. We calculate the dipole matrix element of such transitions and corresponding rate of spontaneous emission for a realistic cavity based on InAlGaAs alloys, investigate its dynamics and estimate quantum efficiency of the terahertz source.
\end{abstract}


\maketitle


\textit{Idea and introduction.---}
It is commonly believed that photons may have only two polarizations (we will refer to them as ``spins" in what follows) corresponding to helicities $\pm 1$ (in units of $\hbar$), being massless particles, or more precisely, massless vector fields \cite{RefMassless}. For a massive field, however, it is always possible to find a so-called ``rest frame" where the spin is zero, which is one of the milestones of the Field theory \cite{RefFieldTheory}. A photon has no rest mass in a free space and therefore for the photon only spin eigenfunction along the direction of propagation exists (lets choose it to be $z$ axis). 
If we introduce a four-vector of electromagnetic potential, ${A_{\mu}}=(\phi,-\mathbf A)$, and the generalised tensor of the electromagnetic field, $F_{\mu\nu}=\partial_\mu A_{\nu}-\partial_\nu A_{\mu}$ for which the Maxwell equations can be written, then the Lagrangian of a massless particles reads ${\cal L}^{M}=-\frac{1}{4}F_{\mu\nu}F^{\mu\nu}$, invariant under local gauge transformation, $A_{\mu} \rightarrow A_{\mu}+\partial_{\mu}\phi_G$.
Thus, the number of the plane wave solutions is reduced to two. Indeed, the Fourier transform of $A_\mu(x)$,
\begin{equation}
\nonumber
A_\mu(k)=\frac{1}{(2\pi)^4}\int A_\mu(x)e^{-ikx}d^4x,
\end{equation}
under a gauge transformation changes as $A_{\mu}(k) \rightarrow A_{\mu}(k)+k_\mu\phi_G$ and the Maxwell equation takes the form
\begin{equation}
\label{EkMaxwellReciprocal}
k^2A_\mu(k)-k_\mu k^\nu A_\nu(k)=0.
\end{equation}
Evidently, $A_\mu(k)$ can be decomposed onto the independent tensors (four-dimensional vectors): (i) $k_\mu=(\mathbf{k},k^0)$; (ii) $p_\mu=(\mathbf{k},-k^0)$; and (iii) $\mathrm{e}_\mu(\mathbf{k},\lambda)$ subject to $k^{\mu}\mathrm{e}_\mu=0$,
\begin{equation}
\nonumber
A_\mu(k)=a^{\lambda}(k)\mathrm{e}_\mu(\mathbf{k},\lambda)+b(k)k_\mu+c(k)p_\mu,
\end{equation}
where $\mathrm{e}_\mu(\mathbf{k},\lambda)$ play a similar role as the unit spinors in the plane-wave decomposition of the Dirac field and they satisfy the orthonormality condition, $\mathrm{e}_\mu(\mathbf{k},\lambda)\mathrm{e}^\mu(\mathbf{k},\lambda')=\delta_{\lambda \lambda'}$. Further, one can find using \eqref{EkMaxwellReciprocal},
\begin{equation}
\nonumber
k^2a^{\lambda}(k)\mathrm{e}_{\mu}(\mathbf{k},\lambda)+c(k)[k^2p_{\mu}-(k\cdot p)k_\mu]=0.
\end{equation}
Since $b(k)$ is absent in this equation, it can be taken arbitrary and thus has no physical meaning. Therefore, two eigen solutions and two corresponding transverse polarization vectors corresponding to two helicities, $\pm 1$ (two spins) are possible.

However, when localized in the travel direction, a photon acquires an effective mass, due to the fact that its dispersion becomes quasiparabolic \cite{RefKlaers,RefLiangCheng},
\begin{equation}
\label{EqPhotonMass}
\hbar \omega=\hbar\frac{c}{n}|\mathbf{k}|=\hbar\frac{c}{n}\sqrt{k_z^2+k_{\parallel}^2}\approx E_0+\frac{\hbar^2k_{\parallel}^2}{2m^*},
\end{equation}
where $z$ is the direction of propagation; $k_{\parallel}^2=k_x^2+k_y^2$; and $m^*$ is $k_z$-dependent effective mass. 
Such a (neutral) massive vector field can be described by the Lagrangian different from the massless case, 
\begin{equation}
\label{EqLagrangian}
{\cal L}^{P}=-\frac{1}{4}F_{\mu\nu}F^{\mu\nu}+\frac{1}{2}m^2A_\nu A^\nu,
\end{equation}
where we introduce the notation $m=(m^*c/\hbar)$ for the normalized mass, and the field equation (which represents a sourceless Proca equation instead of the Maxwell equation) is now \cite{RefProca,RefGreiner}
\begin{equation}
\nonumber
(\Box+m^2)A_{\mu}-\partial_\mu\partial^\nu A_{\nu}=0.
\end{equation}
where $\Box=(-\partial^2/\partial t^2+\partial_\mu\partial^\mu)$ is the d'Alembertian operator.
Similar to the derivation for the massless case above, one can show that here the number of general solutions (representing plane waves) is three,
\begin{equation}
\nonumber
A_{\mu}(x;\mathbf{k},\lambda)=\mathrm{e}_\mu(\mathbf{k},\lambda)\mathrm{exp}[{i\mathbf{k}\cdot\mathbf{x}-i\omega(\mathbf{k})t}],
\end{equation}
where $\omega(\mathbf{k})=c\sqrt{|\mathbf{k}|^2+m^{2}}=k^0/\hbar$. In the basis of independent polarization vectors, $\mathrm{e}_\mu(\mathbf{k},\lambda)$, it is possible to have three polarizations of two different types: two transversal and one longitudinal polarization in the frame $k_\mu=(\mathbf{k},k^0)$:
\begin{eqnarray}
\label{EqPolarizations}
\mathrm{e}_\mu(\mathbf{k},\lambda)&=&(\mathbf{e}^{(\lambda)},0), ~~~
\mathbf{k}\cdot \mathbf{e}^{(\lambda)}=0, ~~~
(\lambda=\pm 1);\\
\nonumber
\mathrm{e}_\mu(\mathbf{k},0)&=&(\mathbf{e}^{(0)},\frac{|\mathbf{k}|}{m}), ~~~ \mathbf{e}^{(0)}=\frac{k^0\mathbf{k}}{m|\mathbf{k}|}.
\end{eqnarray}
A question arises: is it possible to create a photon with zero projection of spin and, even more important, is it feasible to measure it and use?

In this manuscript we propose a setup which can be utilized to create and indirectly detect zero-spin photons and suggest one possible application of these results. For this, we consider a system of exciton polaritons (later ``polaritons") in a semiconductor microcavity. These quasi-particles have hybrid light-matter nature and result from the strong coupling regime in the cavity. Usually, a polariton is formed when a photon interacts with a Wannier-Mott exciton, based on the heavy hole (1S or e1-hh exciton). Due to the fact that the electrons and holes are localized in the semiconductor quantum wells (QWs), the degeneracy between the light hole (lh) and heavy hole branches of the valence band at $k=0$ is lifted. Thus, the energy of the exciton e1-hh is usually lower than the energy of the e1-lh exciton. Therefore, the polaritons based on the excitons e1($\pm 1/2$)-hh($\pm 3/2$) are usually created \cite{RefMicrocavities,RefBaliliScience}. They have finite lifetime since photons leak through the cavity mirrors (DBRs), that is why a constant pumping of the system is required for its operation (see Fig. 1). This pumping, or excitation, can be organized either optically (exposition) \cite{RefKasprzakNature} or electrically (current injection) \cite{RefSchneider2013}. In the latter case, which we will focus on in current Letter, an electron-hole cloud is created; later the carriers of charge form excitons which start to (re)emit and (re)absorb photons in the cavity. This way the exciton polaritons come into play.

It has been shown in a number of theoretical works that such system, in principle, can serve as a source of terahertz (THz) radiation \cite{RefKavokinTHz,RefSavenkoTHz,RefKyriienkoTHz,RefPervishkoOpticsExpress,Todorov3} in the regime of polariton laser generation (spontaneous emission from the quasi-condensate).
THz range still have remained an uncovered region of electromagnetic spectrum due to lack of a solid state source of THz radiation with satisfactory characteristics \cite{Davies,RefOustinov}. 
The main and fundamental objection to creating such a source has been small density of states of THz photons resulting in small rate of spontaneous emission \cite{Duc,Doan}. However fortunately, the emission rate can be increased by application of the Purcell effect if the emitter of THz is placed in a cavity tuned at the THz mode \cite{Todorov1,Chassagneux1}. Moreover, the rate of spontaneous emission of THz photons can be additionally increased by the bosonic stimulation if the radiative transition occurs between the condensate states. For instance, it could be a transition between the upper and lower polariton branches' ground states in the microcavity. However, radiative transition between such modes (originated from the exciton and cavity modes) is forbidden due to the selection rules (since initial and final polariton states correspond to the same exciton and thus have equal parity). The described radiative transition becomes possible if one of the states participating in the photoemission process is hybridized with an exciton state of different parity by an applied electric field \cite{RefKavokinTHz}. In the configuration which we propose in this Letter, the radiative transition can be achieved in a straighter way.


\textit{Theoretical description.---}
We consider a planar quantum microcavity 
in the strong coupling regime, when the exciton polaritons are formed on both the hh and lh excitons. Although this is not a usual situation, the formation of polaritons based on lh(s) has been recently reported experimentally in a number of works (see, for instance, \cite{RefNakayamaPRB,RefAmmerlahnPRB}).
\begin{figure}[!t]
\includegraphics[width=1.0\linewidth]{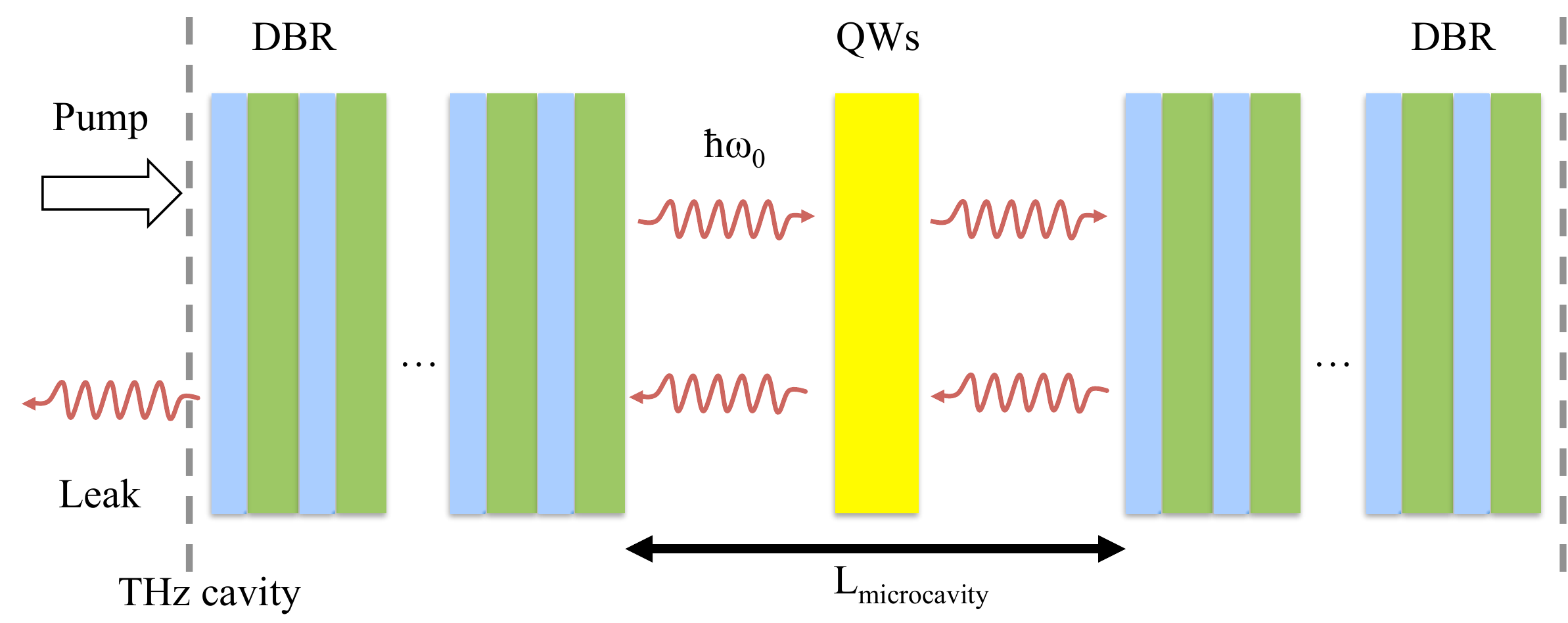}
\caption{Illustration of the system: a semiconductor microcavity under non-resonant excitation (current injection). The photons with frequency $\omega_0$ are localized between two DBRs (distributed Bragg reflectors) and the lh and hh excitons are localised in the QWs (quantum wells). Due to the fact that the excitation is electrical, excitons with zero spin can be created which give birth to the photons with zero spin in the photoabsorption process. An external THz cavity serves to imply the Purcell factor and increase the THz density of states.}
\label{FigSystem}
\end{figure}
We are mostly interested in the subsystem consisting of two lower polariton branches (assuming that scattering from the upper states to the ground states of the dispersions is efficient enough and the particles mostly concentrate on these branches \cite{RefDemenev}), one based on the e1-hh exciton and another one based on the e1-lh exciton. The first (usual) exciton polariton can be formed as a result of interaction between the exciton e1-hh with spin $\pm 3/2\mp 1/2$ and a photon with spin $\pm 1$ due to the selection rules \cite{RefMicrocavities}. Another polariton can have spin projections $\pm 1, 0$ resulted from the interaction of the light hole with spin $\pm 1/2$, an electron $\pm 1/2$, and a photon with spin $\pm 1,0$. 
Accounting for \eqref{EqPolarizations}, the photon field can be described by the vector potential
\begin{eqnarray}
\label{EqCavityVectorPotential}
A_\mu(x)&=&\int d^3k
\sqrt{\frac{\hbar}{2\varepsilon_0\varepsilon\omega(\mathbf{k}) V}}\times\\
\nonumber
&&\sum_{\lambda={0,\pm 1}}
\left(
\mathrm{e}_\mu(\mathbf{k},\lambda)\hat a_{\mathbf{k}\lambda}\mathrm{exp}(i\mathbf{k}\cdot\mathbf{x})
+h.c.
\right),
\end{eqnarray}
where $\hat a_{\mathbf{k}\lambda}$, are the annihilation operators of the photonic modes; $\varepsilon_0$ is the dielectric permittivity of vacuum, $\varepsilon$ is the dielectric constant; and $V$ is the quantization volume. From \eqref{EqCavityVectorPotential} we find the quantized electric field operator in 3D space [in a simplified form: applying the dipole approximation; assuming a single-mode microcavity]:
\begin{eqnarray}
\nonumber
\mathbf{\hat E}
=i\sqrt{\frac{\hbar\omega_0}{2\varepsilon_0\varepsilon V}}\mathbf{e}^{(\lambda)}
(\hat a_C-\hat a_C^\dagger),
\end{eqnarray}
where $\mathbf{{e}}^{(\lambda)}$ is taken from \eqref{EqPolarizations}; $\hat a_C$ is the annihilation operator of the cavity mode; $\omega_0$ is the mode frequency.

The photon with zero spin, obviously, cannot be directly pumped in the cavity externally, that is why optical excitation is of no interest for us. However, an exciton with spin 0 (the result of the union e1 $\pm 1/2$ plus lh $\mp 1/2$) can give birth to such a photon in the photoemission process described by the interaction Hamiltonian,
\begin{eqnarray}
\label{EqInteractionHamiltonian}
\hat{\cal H}_{int}&=&-\mathbf{\hat E}\cdot\mathbf{\hat d}\\
\nonumber
&=&\sqrt{\frac{\hbar\omega_0}{2\varepsilon_0\varepsilon V}}\mathbf{e}^{(\lambda)}
(\hat a_C-\hat a_C^\dagger)
\cdot 
d_0\mathbf{e}_d(\hat a_X-\hat a_X^\dagger),
\end{eqnarray}
where $\hat a_X$ is the annihilation operator of the excitonic mode; $\mathbf{e}_d$ is a unity polarization vector of the dipole; $d_0$ is the amplitude of the dipole matrix elements of the QD exciton transitions which we take equal for simplicity. 
Further, since both $d_0\mathbf{e}_d=(d_x,d_y,d_z)$ and $\mathbf{e}^{(0)}\parallel \mathbf{k}$ may have all the three components non-zero, the matrix element of polariton formation, coming from \eqref{EqInteractionHamiltonian}, is non-zero.
A similar Hamiltonian corresponds to the e1-hh exciton-photon coupling.
Further, we come up with two lower-branch exciton polariton dispersions with spins $\pm 1$ and $\pm 1,0$ and the dispersion relations arising from the coupling determinants
\begin{eqnarray}
\label{EqDeterminant}
\left|\begin{array}{cc}
  \frac{\hbar^2k_{\parallel}^2}{2\mu^{(\alpha)}}-E^{(\alpha)}(\mathbf{k}) & \frac{\hbar\Omega_R}{2}\\
  \frac{\hbar\Omega_R}{2} & \frac{\hbar^2k_{\parallel}^2}{2m^{*}}+\delta^{(\alpha)}-E^{(\alpha)}(\mathbf{k})\\
\end{array}\right|=0,
\end{eqnarray}
where $\alpha=$hh, lh corresponds to two types of excitons; $\mu^{(\alpha)}$ are their effective masses; $\delta^{(\alpha)}$ are the detunings; $E^{(\alpha)}(\mathbf{k})$ are the polariton dispersions; $\Omega_R=2\sqrt{\hbar\omega_0/(2\varepsilon_0\varepsilon V)}d_0/\hbar$ is the Rabi frequency. 
It amounts to the splitting between the upper and lower polariton modes, and in calculations we will set it equal for both the polariton modes, 10 meV, for simplicity. 
It should be noted, that formally the determinant of the system represents a 4 by 4 matrix, which can be split on two 2 by 2 matrices with determinants having form \eqref{EqDeterminant} if we neglect the Coulomb and exchange interaction between the pure excitonic modes.
If we denote the upper state in the system as $|U\rangle=1/\sqrt{2}(|C_U\rangle-|X_U\rangle)$ and the lower state as $|L\rangle=1/\sqrt{2}(|C_L\rangle-|X_L\rangle)$ (see Fig. 2), then the rate of spontaneous emission of THz photons can be estimated from the Fermi golden rule and the Planck formula,
\begin{equation}
\label{EqPlanckFormula}
W= \frac{\omega_{T}\omega_{UL}^2\sqrt{\varepsilon}|d|^2}{3\pi\varepsilon_0\hbar c^3},
\end{equation}
where $\omega_{T}$ is the THz mode frequency; $\omega_{UL}$ is the frequency of the transition from the upper, $|U\rangle$, to the lower, $|L\rangle$, energy state, and we assume $\omega_{T}\approx \omega_{UL}$: the microcavity is embedded into a high quality factor THz cavity with eigen frequency slightly detuned from $\omega_{UL}$ in order to increase the emission efficiency \cite{Chassagneux2,Gallant}; 
the dipole matrix element of the transition accompanied by the emission of the THz photons reads
\begin{equation}
d=e\langle L|\hat r|U\rangle=e\frac{1}{2}\langle X_L|\hat r|X_U\rangle.
\end{equation}
In order to calculate this matrix element, we refer to the analytical eigen solutions for the 2D Hydrogen atom \cite{Ref2DHydrogen} applied to 2D excitons \cite{RefPervishkoOpticsExpress} and find in polar coordinates:
\begin{equation}
\nonumber
d=\frac{e}{2}\int_0^{\infty}\int_0^{2\pi}
\psi_{1l}^{(hh)*}({r},\phi)
r
\psi_{1l}^{(lh)}({r},\phi)
rdrd\phi,
\end{equation}
where 
\begin{eqnarray}
\nonumber
\psi_{nl}^{(\alpha)}({r},\phi)=
\frac{1}{\sqrt{2\pi}}e^{il\phi}
\frac{\beta_n^{(\alpha)}}{(2|l|!)}
\left[\frac{(n-|l|-1)!}{(2n-1)(n-|l|-1)!}\right]^{1/2}\times\\
\nonumber
(\beta_n^{(\alpha)}r)^{|l|}e^{-\frac{\beta_n^{(\alpha)}r}{2}}{_1F_1}(-n+|l|+1,2|l|+1,\beta_n^{(\alpha)}r),
\end{eqnarray}
with $n$ and $l$ being principle and angular momentum quantum numbers, respectively, $n=1,2,3...$, $l=0,\pm 1,\pm 2,\pm (n-1)...$;
\begin{figure}[!t]
\includegraphics[width=1.0\linewidth]{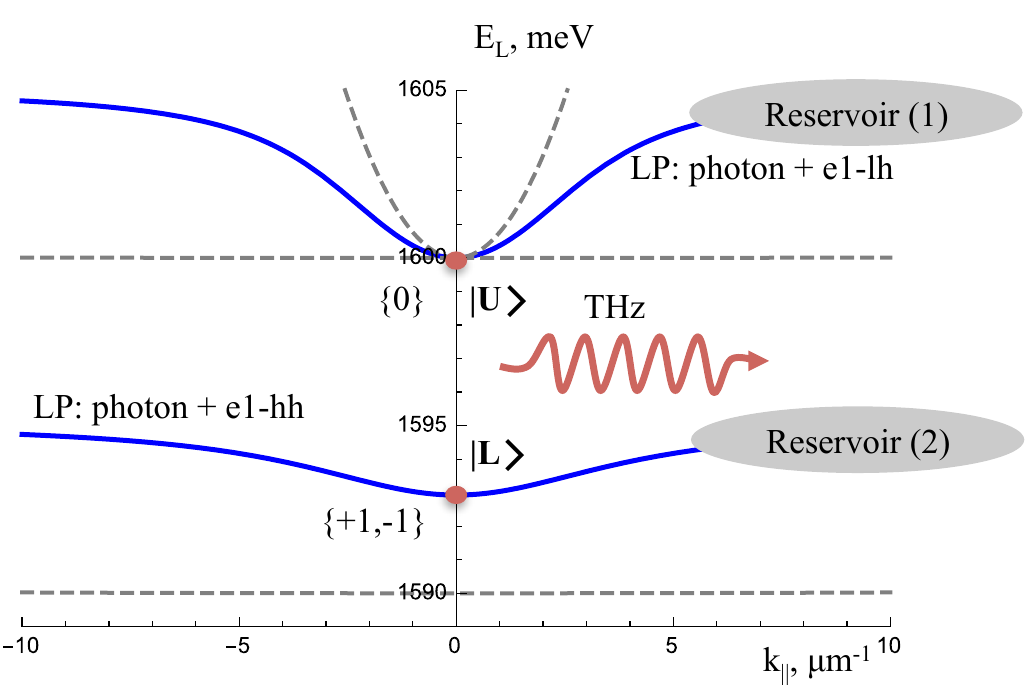}
\caption{Dispersion of the system: energies of particles versus the in-plane pojection of the momentum, $k_{\parallel}$ for the GaAs-based structure. The blue curves correspond to the lower polariton branches for the exciton polaritons based on the e1-hh exciton (lower blue curve) and e1-lh exciton (upper blue curve). The dashed lines show the bare photon and excitons' dispersions. The transition from the upper state $|U\rangle$ with spin $0$ to the lower state $|L\rangle$ with spin $1$ or $-1$ is accompanied by the emission of a THz photon of the corresponding frequency. The polariton modes are also coupled to two incoherent exciton Reservoirs via acoustic phonon-mediated scattering.}
\label{FigDispersion}
\end{figure}
$\beta_n^{(\alpha)}=(2\mu^{(\alpha)} e^2)/(4\pi\varepsilon_0\varepsilon(n-0.5)\hbar^2)$; and ${_1F_1}$ is the confluent hypergeometric function.
In the case $l=0$ which we are interested in, 
\begin{eqnarray}
\nonumber
\psi_{10}^{\alpha}(r,\phi)&=&\frac{1}{\sqrt{2\pi}}\beta_1^{(\alpha)}\mathrm{exp}[{-\frac{\beta_1^{(\alpha)}r}{2}}],
\end{eqnarray}
that yields 
\begin{eqnarray}
\label{EqDipoleElement}
d=e\frac{8\beta_1^{(hh)}\beta_1^{(lh)}}{(\beta_1^{(hh)}+\beta_1^{(lh)})^3}.
\end{eqnarray}
Substituting Eq.\eqref{EqDipoleElement} in \eqref{EqPlanckFormula}, one can calculate the THz emission efficiency for a chosen set of parameters. 

In order to calculate quantum efficiency of this THz source, let us investigate the dynamics of the system using the set of Boltzmann rate equations for the occupation of the upper polariton mode, $n_U$, lower polariton mode, $n_L$, THz mode, $n_T$, and two reservoirs of $k_{\parallel}\neq 0$ states, $n_R^{(1)}$, $n_R^{(2)}$ (see Fig. 2) coupled with the polariton states via acoustic phonons-mediated relaxation:
\begin{eqnarray}
\label{EqBoltzmann}
\dot n_R^{(1)}&=&P-\frac{n_R^{(1)}}{\tau_R^{}}+\frac{n_U-n_R^{(1)}}{\tau_{RU}^{(1)}};\\
\nonumber
\dot n_U&=&-\frac{n_U}{\tau_U}
+\frac{n_R^{(1)}-n_U}{\tau_{RU}^{(1)}}
+\frac{n_R^{(2)}-n_U}{\tau_{RU}^{(2)}}-\tilde W_T;\\
\nonumber
\dot n_R^{(2)}&=&-\frac{n_R^{(2)}}{\tau_R^{}}
+\frac{n_U-n_R^{(2)}}{\tau_{RU}^{(2)}}
+\frac{n_L-n_R^{(2)}}{\tau_{RU}^{(2)}};\\
\nonumber
\dot n_L&=&-\frac{n_L}{\tau_L}
+\frac{n_R^{(2)}-n_L}{\tau_{LU}^{(2)}}+\tilde W_T;\\
\nonumber
\dot n_T&=&-\frac{n_T}{\tau_T}+\tilde W_T,
\end{eqnarray}
$\tilde W_T=W\cdot F_P\left[n_U(n_T+1)(n_L+1)-(n_U+1)n_Tn_L\right]$.
Here $P$ is non-resonant pumping rate; $\tau_R^{}$, $\tau_U$, $\tau_L$, $\tau_T$ are mode lifetimes, whereas $\tau_{RU}^{{(1),(2)}}$, $\tau_{LU}^{(2)}$ are the inverse scattering rates of phonons-assisted transitions. In the steady state ($\dot n_T=\dot n_U=\dot n_L=\dot n_R^{(1)}=\dot n_R^{(2)}=0$), we can estimate the quantum efficiency as $\eta=n_T/(\tau_TP)$.


\textit{Discussions.---}
In a InAlGaAs alloy-based structure with InAs QWs, using \eqref{EqPlanckFormula} we find: $W\approx 5\cdot 10^{-8}$ ps$^{-1}$. It should be noted, that in comparison with the results presented in \cite{RefKavokinTHz} (where the reported rate is $W\approx 10^{-9}$ ps$^{-1}$), the emission rate which we found is more than order of magnitude stronger and can also be additionally stimulated by the Purcell effect. It can be rudimentarily estimated using $W_P=W\cdot F_P$, where $F_P$ may amount to up to 100 \cite{RefKavokinTHz,Todorov1}. 
The external THz cavity can be similar to one discussed in work \cite{RefSavenkoTHz}, therefore we do not present here the details. The only function of this cavity is to increase the rate of spontaneous emission of the THz mode by the Purcell factor, and it does not introduce the qualitative difference to the physical phenomena addressed in current Letter. Besides, since the maximal Purcell factor possible in real setups amounts to dozens and the calculated rate of THz spontaneous emission, $W$, turned out to be more than order of magnitude larger than one in other microcavity-based THz emitters, the requirement of the Purcell cavity can, in principle, be circumscribed which is a valuable advantage of our setup.
Moreover, substituting realistic (quite modest) parameters into Eq.\eqref{EqBoltzmann}: $P=10^3$ ps$^{-1}$, $\tau_U=\tau_L=20$ ps, $\tau_T=10$ ps, $\tau_{RU}^{{(1)}}=\tau_{RU}^{{(2)}}=\tau_{LU}^{(2)}=10$ ps \cite{RefBajoni,RefLevrat}, we find that in the steady state which is established after 150 ps, $\eta\approx 30$\% that sounds very competing.

As mentioned before, the possibility of creating both lh and hh exciton polaritons is justified by recent experimental progress in investigation strained QW structures \cite{RefYoo}. 
The presence of a mechanical pressure, however, is not the only factor which can influence the energy disposition of polariton branches. It is also determined by the Rabi frequency and the detuning between the excitonic and photonic modes (see Fig. 2), beside the semiconductor alloy composition and QWs width. This flexibility in parameters allows us to perform a sort of quantum engineering of the energy structure. The e1-hh polariton can be lower in energy than the e1-lh one, yet it depends on the parameters. (Moreover, in principle, a double-frequency microcavity can be designed with two operational modes tuned at the lh and hh polaritons, correspondingly.) 

The energy difference between the ground states of the polariton modes can vary from fractions of meV to dozens of meV in different alloys corresponding to the THz range. THz photons, resulted from the transitions from the upper state to the lower one, may have spin $\pm 1$. Such photoemission would not only result in indirect (mediated) detection of a photon with zero spin projection, but also serve as a solid-state source of THz radiation with controllable frequency. 
By measuring the THz radiation, it is possible to indirectly address the photons with 0 spin if one manages to either (i) eliminate other possible mechanisms of THz emission; or (ii) tune them at different frequency by the choice of the microcavity geometry and the alloys composition (then the spectral analysis can be applied \cite{RefMuravev}); or (iii) tuning the external THz cavity frequency.
The detector employed to measure this radiation can be of any nature. For instance, it can be an exciton polariton-based detector like one presented in theoretical work \cite{RefSavenkoTHz}, however, it still has not been built experimentally, to the best of our knowledge; or the detector can be of other kind (see e.g. \cite{RefLu,RefMuravev}).


\textit{Conclusions.---} We have considered a system of two cavity exciton-polariton modes coupled by a THz mediated transitions and suggested a method of indirect detection of zero-spin photons in such a microstructure. Besides, we have proposed an alternative source of THz emission based on the transitions between the ground state polariton modes in the regime of polariton Bose-Einstein condensate formation and investigated its dynamics and quantum efficiency with account of the acoustic phonon-mediated relaxation. The emergence of the third (zero) spin projection (third degree of freedom) resulted from the acquisition of the effective mass by the photon localised in the cavity may give birth to a number of interesting phenomena and future applications. 



\end{document}